\documentclass[aps,pra,twocolumn,groupedaddress,showpacs,showkeys]{revtex4-1}
\usepackage[utf8]{inputenc}
\usepackage{amsmath}
\usepackage{amssymb}
\usepackage{graphicx,wrapfig}
\usepackage{pgfplots}
\usepackage[english]{babel}
\usepackage{array}
\usepackage[margin=1in]{geometry}
\usepackage{float}
\usepackage{makecell}
\usepackage{subfig}
\usepackage[colorlinks=true, citecolor=blue]{hyperref}
\usepackage{ragged2e}
\usepackage{multirow}
\usepackage{dcolumn}

\usepackage[none]{hyphenat}

\begin{document}

\title{Automated Error Correction in IBM Quantum Computer and Explicit Generalization}

\author{Debjit Ghosh}
\email{hi2debjitexam@gmail.com}
\affiliation{Indian Institute of Science Education and Research Mohali, Punjab 140306, India}
\author{Pratik Agarwal}
\email{pratikagarwal2203@gmail.com}
\affiliation{Indian Institute of Technology Bombay, Powai, Mumbai, Maharashtra 400076, India}
\author{Pratyush Pandey}
\email{pratyushpandey2001@gmail.com}
\affiliation{University of Notre Dame, Notre Dame, IN 46556, USA}
\author{Bikash K. Behera}
\email{bkb13ms061@iiserkol.ac.in}
\affiliation{Indian Institute of Science Education and Research Kolkata, Mohanpur 741246, India}
\author{Prasanta K. Panigrahi}
\email{panigrahi.iiser@gmail.com}
\affiliation{Indian Institute of Science Education and Research Kolkata, Mohanpur 741246, India}

\begin{abstract}
Experimental realization of automated error correction is demonstrated through IBM Quantum Experience for Bell and GHZ states using a measurement based approach upon ancilla qubits. The measurement automatically activates error correcting unitary operations to restore the system to its original entangled state. We illustrate the algorithm for the maximally entangled qudit case by applying appropriate Hadamard and Controlled-NOT gates.\\
\end{abstract}


\maketitle
\onecolumngrid
\section{Introduction}

Entangled states find a wide range of applications in quantum information processing tasks \cite{nil} and form an integral part of quantum processors. Like highly entangled states, e.g., cluster and Brown \emph{et al.} states, Bell states, GHZ states, and their generalizations have also been extensively used for the implementation of quantum teleportation \cite{ben,tele,sre1,sre3,sre7,sre9}, quantum information splitting \cite{sre2,sre4,sre5,sre7,sre8,sre9,nandi}, quantum secret sharing \cite{sre1,sre3,sre6,sre7,sre10}, superdense coding \cite{sre1,sre9,sup}, quantum cheque \cite{srm,bk1}, and quantum dialogue \cite{dia} to name a few.

These entangled states are prone to partial or complete loss of entanglement due to decoherence, introducing arbitrary phase and bit-flip errors. In particular, as part of a quantum circuit, these subunits need to be monitored for their purity. Appropriate error correction \cite{cor1,cor2,cor3,cor4} needs to be performed so as to render them useful for their assigned tasks in a larger network. A method of non-destructive discrimination of Bell states has been demonstrated by Gupta et al. \cite{gupt1,Gupt}, involving measurement on ancilla qubits, which has subsequently found experimental verification \cite{exp}. This has been extended for automated error correction, wherein the state information on the ancilla is used to recover the assigned state, as and when any error arises due to phase or bit-flip errors \cite{auto}. This state discrimination procedure has been extended for the generalized Bell states as well as for the maximally entangled d-dimensional qudit state \cite{Pani}.

Recently, the Bell state discrimination has been carried out using IBM's cloud based quantum processor \cite{qe}. Being made up of superconducting transmon qubits, it resolves the scalability issues faced by an NMR based quantum computer \cite{Jhar,anil}. This has opened the avenue for checking the efficacy of several algorithms and protocols, e.g., researchers have been successfully implemented a number of applications in quantum information theory using IBM quantum computer \cite{exp,IBM1,IBM2,bk1,IBM4,IBM5}. Here, we demonstrate the automated error correction for the Bell states. Subsequently, the GHZ state is discriminated non-destructively and the corresponding error correction is experimentally demonstrated. The quantum state tomography is carried out with the associated simulation to explicate the efficacy of our method. We then proceed to give complete error correction algorithm for the maximally entangled qudit case.

The paper is organized as follows. In section-\ref{secnon}, we have described non-destructive discrimination (both theoretical and experimental) of GHZ state. In section-\ref{secauto}, we have experimentally verified the automated error correction algorithm for Bell and GHZ states in the IBM quantum computer. Section-\ref{sectomo} deals with the state tomography of our experimental circuits. We generalize the automated error correction algorithm for n-qudits in section-\ref{secqudit}. We conclude in section-\ref{seccon} by providing future directions of this work. Results of all experimental circuits and calibration data of the IBM quantum processor are given in appendix-\hyperlink{ap1}{1} and appendix-\hyperlink{ap2}{2} respectively.

\section{Non-destructive discrimination algorithm}\label{secnon}
Generalization of non-destructive discrimination algorithm has been proposed by Gupta et al. \cite{gupt1}. $|\psi_x\rangle$ is the generalized Bell states (GBS) and the initial state of any ancilla (represented as $|a\rangle$ or $|a_i\rangle$) is $|0\rangle$. The final state of the ancilla represents the phase, as $|\phi\rangle$ or the parity, represented as $|p_i\rangle$ (where $i\in[1,n-1]$). It is to be noted that, $|\phi\rangle = |0\rangle$ or $|1\rangle$ corresponds to `+' or `-' phase respectively. 
\subsection{Theoretical protocol}
\begin{itemize}
	\item{\textbf{Phase checking:}} 
	\begin{figure}[H]
		\centering
		\includegraphics[scale=0.4]{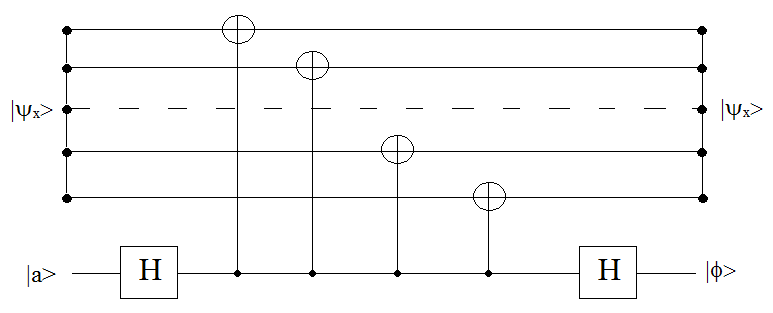}
		\caption{\label{phcheck}\emph{Circuit illustrating phase of the ancilla.}}
	\end{figure}

\item{\textbf{Parity checking:}} 
\begin{figure}[H]
	\centering
	\includegraphics[scale=0.4]{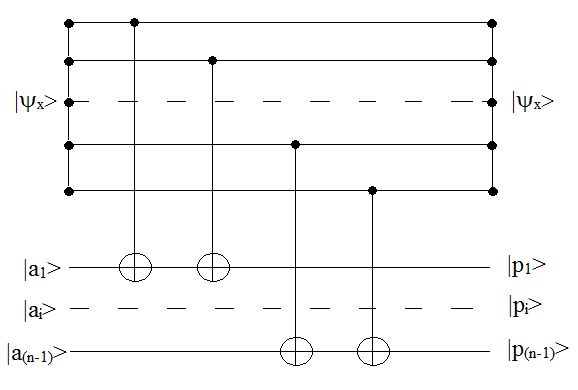}
	\caption{\label{prcheck}\emph{Circuit illustrating parity of the ancilla.}}
\end{figure}
\end{itemize}

\begin{table}[h]
\centering
\caption{\emph{The measurement results corresponding to different Bell and GHZ states are depicted in the following table.}}
\begin{tabular} {|c|c|c|c|c|c|}
\hline
\mbox{Bell States} & $|p\rangle$ & $|\phi\rangle$ & \mbox{GHZ States} & $|p\rangle$ & $|\phi\rangle$\\
\hline
$|\Psi_{00}^{+}\rangle$ = $\frac{1}{\sqrt{2}}(|00\rangle + |11\rangle)$ & $|0\rangle$ & $|0\rangle$ & $|\Psi_{000}^\pm\rangle$ = $ \frac{1}{\sqrt{2}}(|000\rangle \pm |111\rangle)$ & $|00\rangle $ & $|0\rangle/|1\rangle$\\

$|\Psi_{00}^{-}\rangle$ = $\frac{1}{\sqrt{2}}(|00\rangle - |11\rangle)$ & $|0\rangle$ & $|1\rangle$ & $|\Psi_{001}^\pm\rangle = \frac{1}{\sqrt{2}}(|001\rangle \pm |110\rangle)$ & $|01\rangle$ & $|0\rangle/|1\rangle$\\

$|\Psi_{01}^{+}\rangle = \frac{1}{\sqrt{2}}(|01\rangle + |10\rangle)$ & $|1\rangle$ & $|0\rangle$ & $|\Psi_{010}^\pm\rangle = \frac{1}{\sqrt{2}}(|010\rangle \pm |101\rangle)$ & $|11\rangle$ & $|0\rangle/|1\rangle$\\

$|\Psi_{01}^{-}\rangle = \frac{1}{\sqrt{2}}(|01\rangle - |10\rangle)$ & $|1\rangle$ & $|1\rangle$ & $|\Psi_{011}^\pm\rangle = \frac{1}{\sqrt{2}}(|011\rangle \pm |100\rangle)$ & $|10\rangle$ & $|0\rangle/|1\rangle$\\
\hline
\end{tabular}
\end{table}
\subsection{Experimentally verified with GHZ states}\label{seccheck}
Quantum circuits and methods, used for nondestructive discrimination of Bell states, have been experimentally realized using IBM quantum computer \cite{exp}. Due to the lack of coupling with all the qubits, CNOT gate is not accessible to all the qubits. Hence, the process of qubit swapping \cite{user} can be used to overcome this restriction. In the correction algorithm (phase flip correction), controlled Z gate is used to correct the phase of the erroneous state. An equivalent quantum circuit for controlled Z gate (composed of CNOT and X gates) can be found in the user manual \cite{user}. 

\begin{figure}[H]
\includegraphics[scale=0.3]{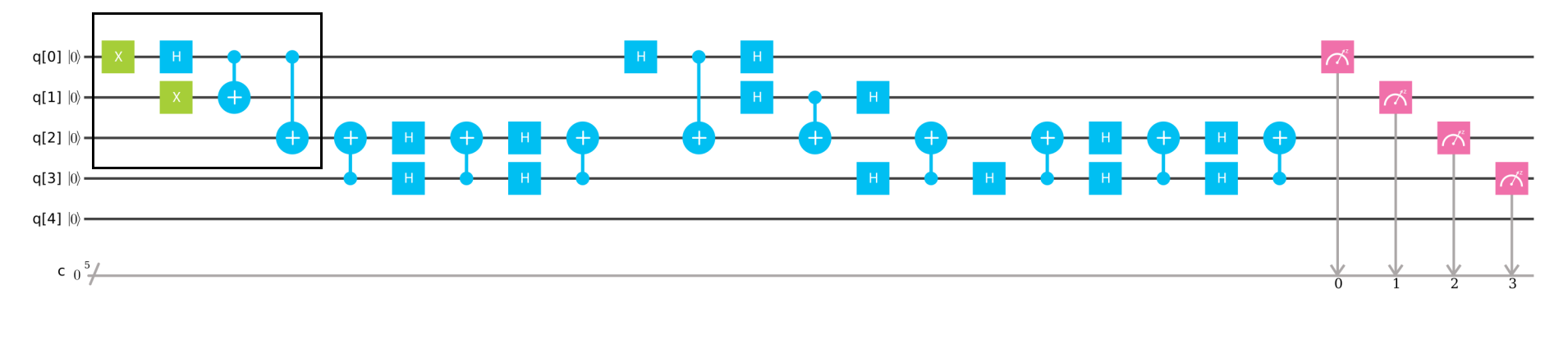}
\caption{\label{phc}Phase-checking circuit in IBM quantum computer. The box part generates the following GHZ state, $|\Psi_{010}^{-}\rangle$.}
\end{figure}

The phase of $|\Psi_{010}^{-}\rangle$ is $|1\rangle$ (results are given in \hyperlink{ap1}{Appendix 1}). After constructing all GHZ states, the phase of ancilla for all $|\Psi_x^{+}\rangle$ and $|\Psi_x^{-}\rangle$ are observed to be $|0\rangle$ and $|1\rangle$ respectively.
\begin{figure}[H]
\includegraphics[scale=0.25]{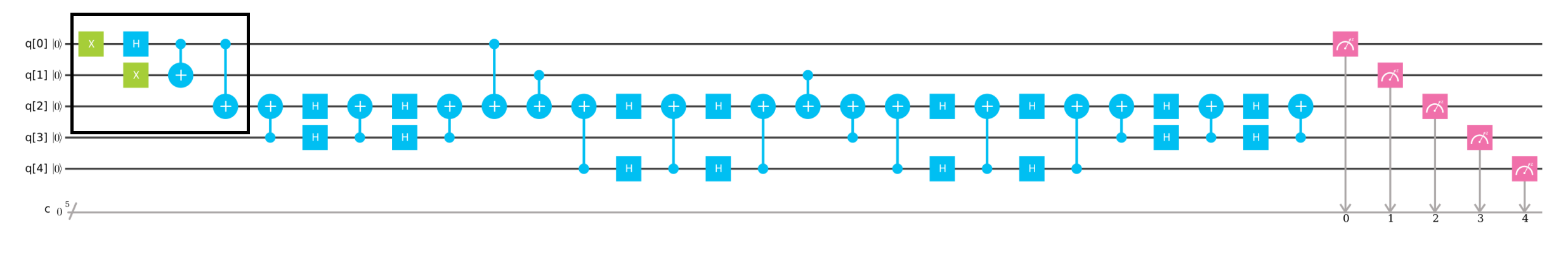}
\caption{\label{prc}Parity-checking circuit in IBM quantum computer. The box part generates the following GHZ state, $|\Psi_{010}^{-}\rangle$.}
\end{figure}
The parity of $|\Psi_{010}^{-}\rangle$ is $|11\rangle$ (results are given in \hyperlink{ap1}{Appendix 1}). The parity of all GHZ states is checked by changing the initial gates inside the box, shown in the above figure. 

\section{Automated error correction algorithm}\label{secauto}
Automated error correction algorithm \cite{auto} requires two steps to eliminate arbitrary phase change error and bit-flip error. Arbitrary phase change error also requires two steps, which include removing arbitrary phase errors followed by a phase flip operation \cite{auto}.  
\subsection{Theoretical protocol}
Any erroneous state is represented by superscript \textquoteleft ek\textquoteright, where k$\in$\{0,1,2,...\} and final ancilla is denoted by the superscript \textquoteleft f\textquoteright. 
\newpage
\begin{itemize}
	\item {\bfseries Step 1:} Arbitrary phase error correction
\begin{figure}[H]
	\centering
	\includegraphics[scale=0.4]{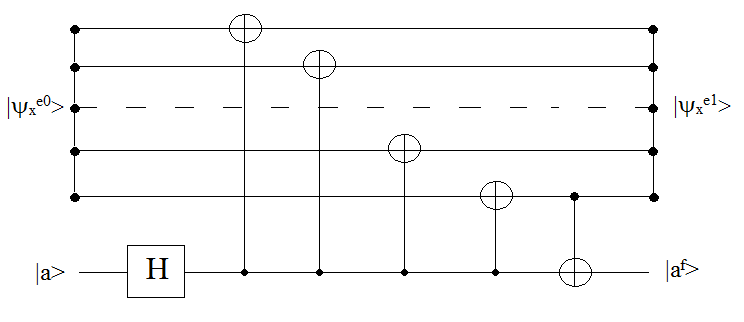}
	\caption{\label{error}Circuit depicting arbitrary phase error correction.}
\end{figure}
	\item {\bfseries Step 2:} Phase-flip error correction
\begin{figure}[H]
\centering
\includegraphics[scale=0.4]{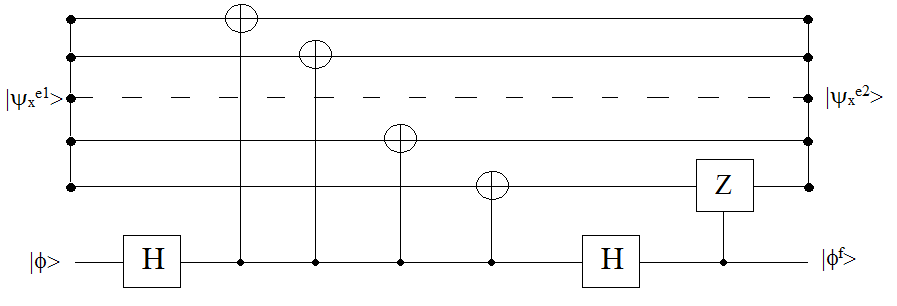}
\caption{\label{phcor}Circuit depicting phase-flip error correction.}
\end{figure}
	\item {\bfseries Step 3:} Bit-flip error correction
\begin{figure}[H]
	\centering
	\includegraphics[scale=0.5]{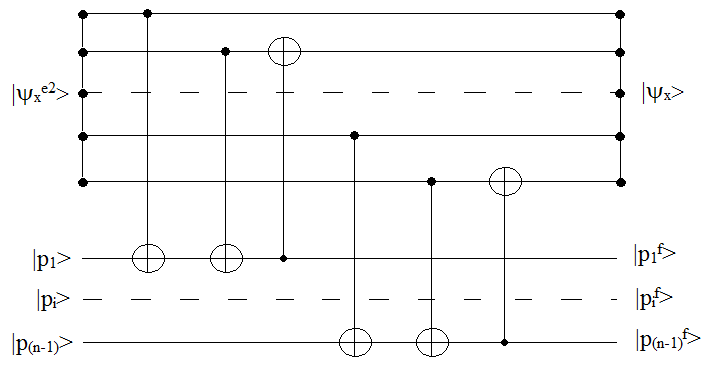}
	\caption{\label{prcor}Circuit depicting bit-flip error correction.}
\end{figure}
\end{itemize}

The circuit presented in fig-\ref{error}, removes arbitrary phase from the erroneous state. Thus only phase-flip error remains in the erroneous state which is corrected in the next step. After Step 1, the phase of the erroneous state is $|0\rangle$. In step 2, the circuit shown in fig-\ref{phcor}, flips the phase of the erroneous state to its correct initial phase $|\phi\rangle$. Finally, circuit in fig-\ref{prcor}, flips the bit of parity of erroneous state to its correct initial bit of the parity $|p_i\rangle$.

\subsection{Experimentally verified with Bell states}\label{secbell}
We construct the automated error correction algorithm\cite{auto} in IBM quantum computer by using the above equivalent circuits. 

\begin{figure}[H]	
	\includegraphics[height=4 cm,width=18cm]{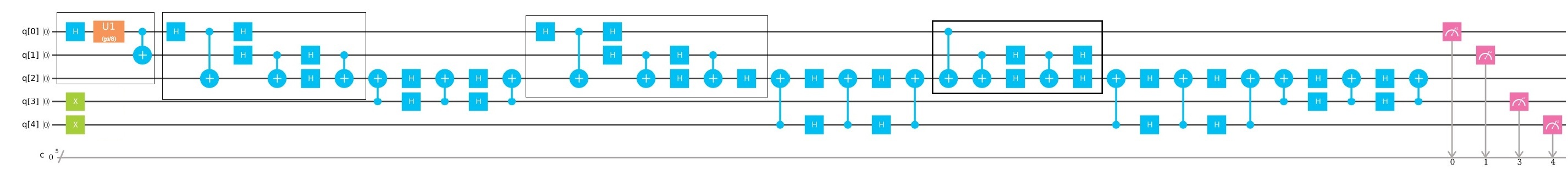}
	\caption{\label{bell}Error correction of Bell state is depicted in the above circuit. First box is used to create a Bell state with arbitrary phase error, phase flip error and bit flip error. Second box is used for arbitrary phase correction, which is followed by the third and the last box demonstrating both phase flip and bit flip correction respectively. The orange box in the first box represents a $\pi/8$ phase shift gate.}
	
\end{figure}

Ancillas of arbitrary phase correction, phase check and parity check are q[2], q[3] and q[4] respectively (fig-\ref{bell}). The initial phase and parity are given correspondingly, $|\phi\rangle=|1\rangle$, and $|p\rangle=|1\rangle$. This circuit corrects $|\Psi^e\rangle=\frac{1}{\sqrt{2}}(|00\rangle+e^{\iota\pi/8}|11\rangle)$ to $|\Psi\rangle=\frac{1}{\sqrt{2}}(|01\rangle-|10\rangle)$ (results are given in \hyperlink{ap1}{Appendix 1}). Correction of other Bell states can be done by changing the state of parity and phase in the same circuit.
When there is no error in phase, the final state of phase will be $|0\rangle$. But if there is a flip error in initial state with respect to initial phase then the final state of phase will be $|1\rangle$. This also holds same for bit flip error correction.

\subsection{Experimentally verified with GHZ states}\label{seccor}
According to the algorithm, correction should be done in three steps \cite{auto}. Due to 5 qubit quantum computer, all three steps can't be done in one circuit just like Bell state(fig-\ref{bell}), as 6 qubits are needed (three qubits for GHZ state, one qubit for arbitrary phase correction, one qubit for phase-flip correction and two qubits for bit-flip correction). Hence, the following steps need to be carried out as shown in the below three circuits. 

\subsubsection{Arbitrary phase correction}
For GHZ states, the same algorithm is constructed as in case of Bell states. 
The correction of an erroneous state $|\Psi^e\rangle=\frac{1}{\sqrt{2}}(|000\rangle+e^{\iota\pi/8}|111\rangle)$ for initial phase $|\phi\rangle=|1\rangle$ and initial parity $|p\rangle=|11\rangle$ is described below.\\
This step is concerned with removing arbitrary phase (here $e^{\iota\pi/8}$), which further has a phase-flip error correction, from initial erroneous state as in the case of the second box of the Bell state correction (fig-\ref{bell}).

\begin{figure}[H]
\begin{center}
\includegraphics[scale=0.35]{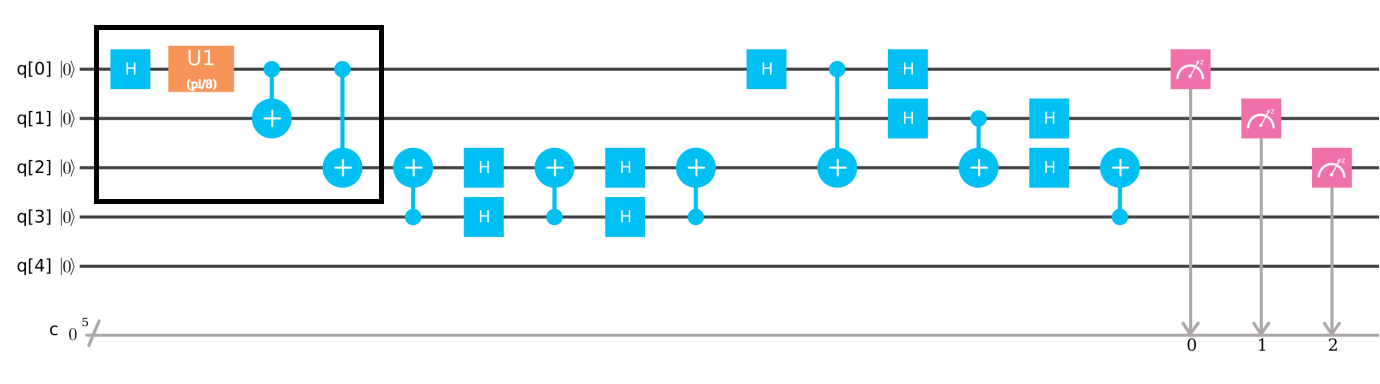}
\end{center}
\caption{\label{phe}Circuit depicting arbitrary phase correction. The box part creates $|\Psi^{e}\rangle$ state, where the orange box is represented for the $\pi/8$ phase shift gate.}
\end{figure}

This circuit (fig-\ref{phe}) converts the state $|\Psi^{e}\rangle$ into $|\Psi_{000}\rangle=\frac{1}{\sqrt{2}}(|000\rangle+|111\rangle)$ (results are given in \hyperlink{ap1}{Appendix 1}).
\subsubsection{Phase flip correction}
Arbitrary phase has been removed. Now this step deals with the correction of phase flip error, same as the third box of Bell states correction (fig-\ref{bell}). This circuit follows the erroneous state as phase checking, but in the end, if there exits a flip then a controlled Z gate comes into play to fix the flip error.

\begin{figure}[H]
	\begin{center}
		\includegraphics[scale=0.27]{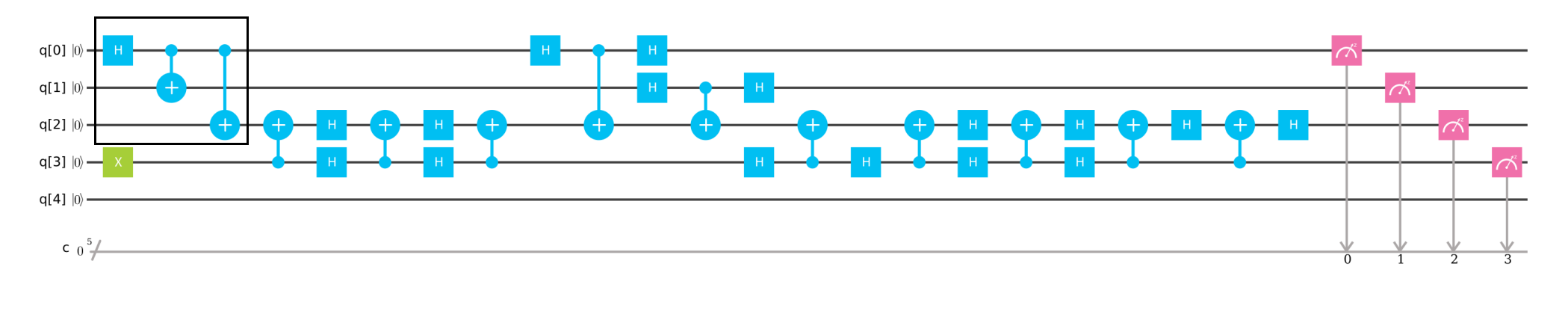}
	\end{center}
	
	\caption{\label{phase}Circuit illustrating phase flip correction. The box part creates the state, $|\Psi_{000}^{+}\rangle$, where initial phase, $|\phi\rangle=|1\rangle$.}
\end{figure} 

After the application of the circuit (fig-\ref{phase}) phase of the state change from (+) to (-). The state $|\Psi_{000}^{+}\rangle$ will change to $|\Psi_{000}^{-}\rangle$ (results are given in \hyperlink{ap1}{Appendix 1} ) as the initial phase $|\phi\rangle=|1\rangle$. The final state of $|\phi\rangle$ will be $|1\rangle$ as it changes the phase of the erroneous state.

\subsubsection{Bit flip correction}\label{secbit}
The last step of the algorithm is the bit-flip correction, same as the last box of Bell states correction (fig-\ref{bell}). Just like the phase correction, the circuit follows erroneous state as parity checking and flip the parity of the state by CNOT gate if it doesn't match with initial parity.

\begin{figure}[H]
	\begin{center}
		\includegraphics[scale=0.21]{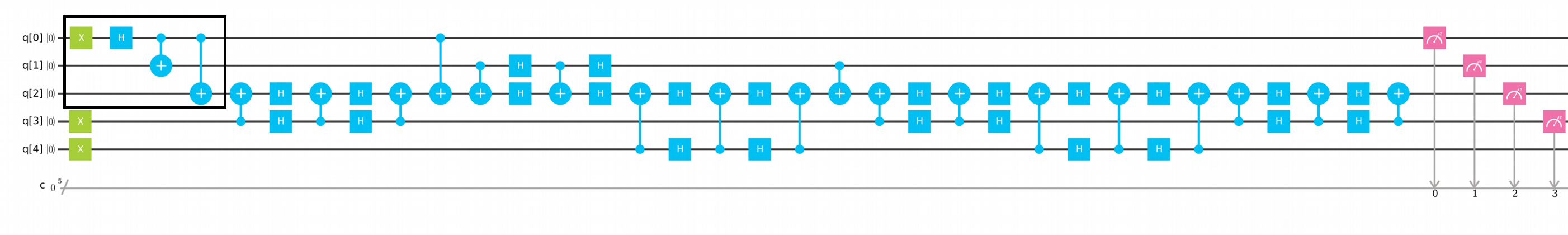}
	\end{center}
	
	\caption{\label{parity}Circuit illustrating bit flip correction. The box part creates the state, $|\Psi_{000}^{-}\rangle$, where initial parity, $|p\rangle=|11\rangle$.}
\end{figure} 

After the implementation of the circuit (fig-\ref{parity}) parity will be corrected i.e. the state $|\Psi_{000}^{-}\rangle$ will change to our desired or correct state $|\Psi_{010}^{-}\rangle$ (results are given in \hyperlink{ap1}{Appendix 1}). The final state of parity after correction will be $|10\rangle$. From the final state of parity after correction we can have the information about the GHZ erroneous state after arbitrary correction. This information is given in the following table.\\

\begin{table}[H]
	\begin{center}

		\begin{tabular}{|c|c|c|c|c|c|}
			\hline 
			State & \thead{ Parity \\check \\result} & \thead{Final parity \\state for initial \\$parity=|00\rangle$} & \thead{Final parity \\state for initial\\ $parity=|01\rangle$} & \thead{Final parity \\state for initial \\$parity=|10\rangle$} & \thead{Final parity \\state for initial \\$parity=|11\rangle$} \\ 
			\hline 
			$|\Psi_{000}^\pm\rangle$ & $|00\rangle$ & $|00\rangle$ & $|01\rangle$ & $|11\rangle$ & $|10\rangle$ \\ 
			\hline 
			$|\Psi_{001}^\pm\rangle$ & $|01\rangle$ & $|01\rangle$ & $|00\rangle$ & $|10\rangle$ & $|11\rangle$ \\ 
			\hline 
			$|\Psi_{010}^\pm\rangle$ & $|11\rangle$ & $|10\rangle$ & $|11\rangle$ & $|01\rangle$ & $|00\rangle$ \\ 
			\hline 
			$|\Psi_{011}^\pm\rangle$ & $|10\rangle$ & $|11\rangle$ & $|10\rangle$ & $|00\rangle$ & $|01\rangle$ \\ 
			\hline 
		\end{tabular}
	\end{center}
	
\end{table}

Corrections to other GHZ states can be done by changing the desire state of initial phase and parity. 
\section{Quantum state tomography}\label{sectomo}
Quantum state tomography is performed to calculate an experimental density matrix which helps to check the accuracy of the experimental process. Experimental density matrix of n qubits is, 
\begin{equation}
\rho^{E}=\frac{1}{2^n}\sum_{i_1,i_2,...,i_n=0}^{3}\langle\sigma_{i_1}\otimes\sigma_{i_2}\otimes...\otimes\sigma_{i_n}\rangle(\sigma_{i_1}\otimes\sigma_{i_2}\otimes...\otimes\sigma_{i_n})
\end{equation}
where $i_k$, $k\in(1,2,...,n)$, $\sigma_0=I$, $\sigma_1=X$, $\sigma_2=Y$ and $\sigma_3=Z$ Pauli matrices.
$\langle\sigma_{i_1}\otimes\sigma_{i_2}\otimes...\otimes\sigma_{i_n}\rangle$ is evaluated experimentally by obtaining the respective probabilities when the qubits are measured in standard basis (\cite{14,26,27,28,29,30,31}). IBM Experience uses Z-basis measurement. A Hadamard gate (H) is added before measurement for X-basis measurement and the phase shift gate $S^{\dagger}$ before Hadamard gate is added for Y-basis measurement. We build experimental density matrices of the required final state of the above circuits. Each measurement runs 8192 times. Each density matrix is composed of a real part(Re[$\rho^E$]) and imaginary part(Im[$\rho^E$]). We have plotted the real part of each experimental density matrix with its theoretical density matrix. After constructing density matrices, fidelity can be obtained as a quantitative measure of accuracy. Fidelity, $F(\rho^T,\rho^E)$ = Trace$(\sqrt{\sqrt{\rho^T}.\rho^E.\sqrt{\rho^T}})$ \cite{32} where $\rho^T$ is the theoretical density matrix. Since all our theoretical density matrices are pure i.e. Trace($\rho^2$)$=1$, we have used another definition of fidelity, $F(\rho^T,\rho^E)=\sqrt{\langle\Psi|\rho^E|\Psi\rangle}$ \cite{32}, where $\rho^T=|\Psi\rangle\langle\Psi|$. Average absolute deviation, $\langle\Delta x\rangle=\frac{1}{n^2}\sum_{i,j=1}^{n}|x^{T}_{i,j}-x^{E}_{i,j}|$ and maximum absolute deviation, $\Delta x_{max}=Max|x^{T}_{i,j}-x^{E}_{i,j}|$, where $i,j \in[1,n]$ and $x^{T}_{i,j}$ and $x^{E}_{i,j}$ are the $i^{th}$ row and $j^{th}$ column elements of $\rho^{T}$ and $\rho^{E}$ respectively. 

\subsection{Phase checking}

\subsubsection{GHZ state}
In section-\ref{seccheck} phase checking circuit is constructed for a state $|\Psi^{-}_{010}\rangle$. The GHZ state $|\Psi^{-}_{010}\rangle$ must be verified for non-destructive discrimination of that state. First three qubits which are encoded for GHZ state, are required to do complete state tomography to construct the density matrix.\\
The theoretical density matrix,
\begin{equation}
\begin{split}
\rho^T&=|\Psi^{-}_{010}\rangle\langle\Psi^{-}_{010}|\\
&=\begin{pmatrix}
0 & 0 & 0 & 0 & 0 & 0 & 0 & 0\\
0 & 0 & 0 & 0 & 0 & 0 & 0 & 0\\
0 & 0 & 0.5 & 0 & 0 & 0.5 & 0 & 0\\
0 & 0 & 0 & 0 & 0 & 0 & 0 & 0\\
0 & 0 & 0 & 0 & 0 & 0 & 0 & 0\\
0 & 0 & 0.5 & 0 & 0 & 0.5 & 0 & 0\\
0 & 0 & 0 & 0 & 0 & 0 & 0 & 0\\
0 & 0 & 0 & 0 & 0 & 0 & 0 & 0\\
\end{pmatrix}
\end{split}
\end{equation}
The experimental density matrix,
\begin{equation*}
\rho^E= Re[\rho^E] + \iota Im[\rho^E]
\end{equation*} 

\begin{equation}
Re[\rho^E]=
\begin{pmatrix}
0.0000 &  -0.0016  &  0.0019 &  -0.0024  &  0.0019  & -0.0040 &   0.0030     &    0\\
-0.0016  &  0.0000  & -0.0001 &  -0.0059  &  0.0018  & -0.0051    &     0  & -0.0020\\
0.0019  & -0.0001   & 0.4940   &-0.0016  &  0.0008  & -0.5000  &  0.0009  & -0.0005\\
-0.0024  & -0.0059 &  -0.0016  & 0.0000    &     0  &  0.0003  &  0.0057  &  0.0074\\
0.0019  &  0.0018   & 0.0008     &    0   & 0.0000  & -0.0026   & 0.0004  & -0.0031\\
-0.0040  & -0.0051 &  -0.5000  &  0.0003 &  -0.0026  &  0.5060  &  0.0006  &  0.0006\\
0.0030    &     0  &  0.0009  &  0.0057 &   0.0004  &  0.0006 &  0.0000  &  0.0039\\
0  & -0.0020  & -0.0005  &  0.0074 &  -0.0031  &  0.0006   & 0.0039   & 0.0000\\
\end{pmatrix}
\end{equation}
\begin{equation}
Im[\rho^E]=
\begin{pmatrix}
         0  & -0.0049 &  -0.0029 &  -0.0025  & -0.0008  &  0.0011  &       0  &  0.0080\\
0.0049      &   0  &  0.0025 &  -0.0029  & -0.0006 &  -0.0030  &  0.0020     &    0\\
0.0029  & -0.0025   &      0  &  0.0054  & -0.0027  &  0.0030 &   0.0065 &  -0.0011\\
0.0025  &  0.0029  & -0.0054  &       0  & -0.0020   & 0.0037  & -0.0014  &  0.0023\\
0.0008   & 0.0006  &  0.0027  &  0.0020      &   0  & -0.0004 &  -0.0024  &  0.0005\\
-0.0011   & 0.0030 &  -0.0030 &  -0.0037 &   0.0004    &     0 &  -0.0015 &   0.0031\\
0 &  -0.0020 &  -0.0065  &  0.0014 &   0.0024 &   0.0015   &      0  &  0.0029\\
-0.0080     &    0  &  0.0011 &  -0.0023 &  -0.0005 &  -0.0031 &  -0.0029  &       0
\end{pmatrix}
\end{equation}
\begin{figure}[H]
	\begin{center}
		\includegraphics[scale=0.6]{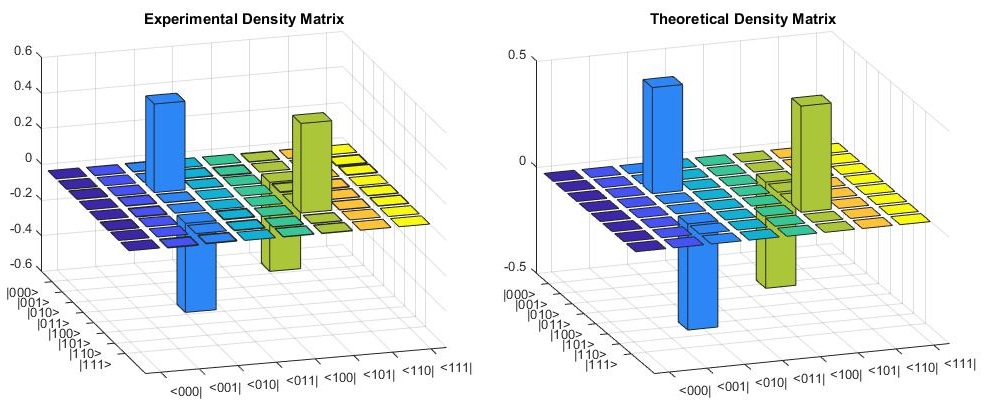}
		\caption{\label{t_phc}Construction of GHZ state $|\Psi^{-}_{010}\rangle$.}
	\end{center}
\end{figure}

Average absolute deviation, $\langle\Delta x\rangle=0.2\% $, maximum absolute deviation, $\Delta x_{max}=0.74\%$ and fidelity, $F(\rho^T,\rho^E)=\sqrt{\langle\Psi^{-}_{010}|\rho^E|\Psi^{-}_{010}\rangle}= 1$.
\subsubsection{Ancilla}
As the state $|\Psi^{-}_{010}\rangle$ is used for the circuit, the ancilla has to be $|\Psi\rangle=|1\rangle$. Ancilla is encoded into q[3] of the circuits, which means a complete state tomography is done on this qubit.\\
The theoretical density matrix,
\begin{equation}
\begin{split}
\rho^T&=|\Psi\rangle\langle\Psi|\\
&=\begin{pmatrix}
     0 &    0\\
0    & 1
\end{pmatrix}
\end{split}
\end{equation}
The experimental density matrix,
\begin{equation*}
\rho^E= Re[\rho^E] + \iota Im[\rho^E]
\end{equation*} 
\begin{equation}
Re[\rho^E]=
\begin{pmatrix}
   0  &  0.0010\\
0.0010  &  1.0000
\end{pmatrix}
\end{equation}
\begin{equation}
Im[\rho^E]=
\begin{pmatrix}
   0  &  0.0010\\
-0.0010  &  0
\end{pmatrix}
\end{equation}
\begin{figure}[H]
	\begin{center}
		\includegraphics[scale=0.5]{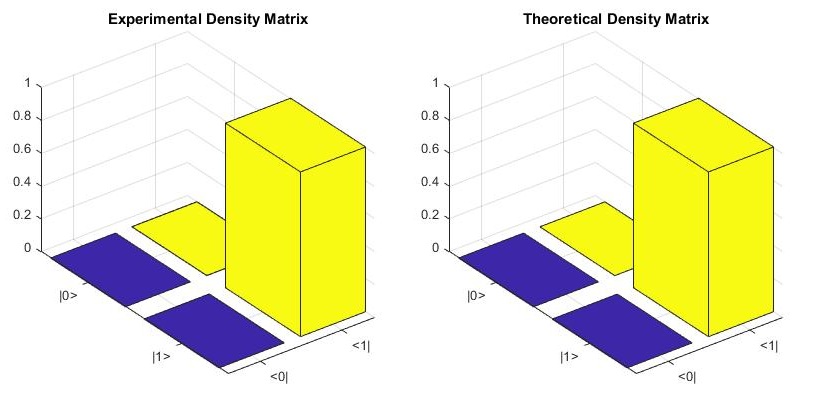}
		\caption{\label{t_phasean}Construction of the ancilla $|\Psi\rangle=|1\rangle$.}
	\end{center}
\end{figure}
Average absolute deviation, $\langle\Delta x\rangle=0.05\% $, maximum absolute deviation, $\Delta x_{max}=0.1\%$ and fidelity, $F(\rho^T,\rho^E)=\sqrt{\langle\Psi|\rho^E|\Psi\rangle}= 1$.

\subsection{Parity checking}

\subsubsection{GHZ state}
In section-\ref{seccheck} parity checking circuit describes the parity of the GHZ state $|\Psi^{-}_{010}\rangle$. A complete state tomography of first three qubits(q[0], q[1], q[2]) which are encoded for GHZ state in the circuit, is needed to verify the non-destructive discrimination of the state $|\Psi^{-}_{010}\rangle$.\\
The theoretical density matrix,
\begin{equation}
\begin{split}
\rho^T&=|\Psi^{-}_{010}\rangle\langle\Psi^{-}_{010}|\\
&=\begin{pmatrix}
0 & 0 & 0 & 0 & 0 & 0 & 0 & 0\\
0 & 0 & 0 & 0 & 0 & 0 & 0 & 0\\
0 & 0 & 0.5 & 0 & 0 & 0.5 & 0 & 0\\
0 & 0 & 0 & 0 & 0 & 0 & 0 & 0\\
0 & 0 & 0 & 0 & 0 & 0 & 0 & 0\\
0 & 0 & 0.5 & 0 & 0 & 0.5 & 0 & 0\\
0 & 0 & 0 & 0 & 0 & 0 & 0 & 0\\
0 & 0 & 0 & 0 & 0 & 0 & 0 & 0\\
\end{pmatrix}
\end{split}
\end{equation}
The experimental density matrix,
\begin{equation*}
\rho^E= Re[\rho^E] + \iota Im[\rho^E]
\end{equation*} 

\begin{equation}
Re[\rho^E]=
\begin{pmatrix}
   0.0000 &  -0.0013 &   0.0024 &  -0.0004 &   0.0001 &  -0.0025  & -0.0004  & -0.0002\\
-0.0013  &  0.0000  &  0.0011  &  0.0029  & -0.0040 &  -0.0006 &   0.0002 &  -0.0041\\
0.0024  &  0.0011   & 0.4970  &  0.0018  & -0.0009 &  -0.5000  & -0.0004  &  0.0040\\
-0.0004  &  0.0029 &   0.0018  & 0.0000 &  0.0000  &  0.0044  & -0.0035 &  -0.0001\\
0.0001  & -0.0040 &  -0.0009 &  0.0000  &  0.0000  &  0.0008 &  -0.0016  &  0.0024\\
-0.0025  & -0.0006  & -0.5000  &  0.0044  &  0.0008  &  0.5030  & -0.0021 &   0.0024\\
-0.0004   & 0.0002 &  -0.0004  & -0.0035  & -0.0016  & -0.0021 &  0.0000 &  -0.0003\\
-0.0002 &  -0.0041  &  0.0040  & -0.0001 &   0.0024  &  0.0024  & -0.0003  &  0.0000
\end{pmatrix}
\end{equation}
\begin{equation}
Im[\rho^E]=
\begin{pmatrix}
         0 &   0.0008  &  0.0044 &  -0.0009 &   0.0037  &  0.0041 &  -0.0036 &  -0.0026\\
-0.0008    &     0  & -0.0056  &  0.0039   & 0.0016  &  0.0017 &  -0.0036  & -0.0061\\
-0.0044  &  0.0056   &      0  &  0.0053  &  0.0041  &  0.0009  &  0.0027 &  -0.0021\\
0.0009 &  -0.0039  & -0.0053    &     0  &  0.0019  &  0.0001  & -0.0006   & 0.0007\\
-0.0037 &  -0.0016 &  -0.0041  & -0.0019   &      0  &  0.0023 &  -0.0006 &  -0.0021\\
-0.0041  & -0.0017 &  -0.0009 &  -0.0001  & -0.0023   &      0  & -0.0004  &  0.0074\\
0.0036  &  0.0036  & -0.0027  &  0.0006  &  0.0006  &  0.0004  &       0  &  0.0018\\
0.0026  &  0.0061 &   0.0021  & -0.0007 &   0.0021 &  -0.0074 &  -0.0018   &      0
\end{pmatrix}
\end{equation}
\begin{figure}[H]
	\begin{center}
		\includegraphics[scale=0.6]{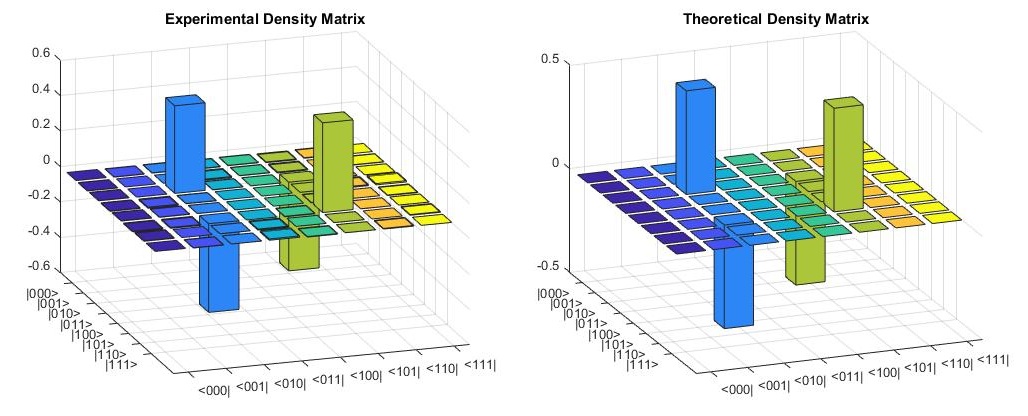}
		\caption{\label{t_prc}Construction of the GHZ state $|\Psi^{-}_{010}\rangle$}.
	\end{center}
\end{figure}

Average absolute deviation, $\langle\Delta x\rangle=0.15\% $, maximum absolute deviation, $\Delta x_{max}=0.44\%$ and fidelity, $F(\rho^T,\rho^E)=\sqrt{\langle\Psi^{-}_{010}|\rho^E|\Psi^{-}_{010}\rangle}= 1$.

\subsubsection{Ancilla}
Parity of the required state $|\Psi^{-}_{010}\rangle$ is $|\Psi\rangle=|11\rangle$. Parity of the state is encoded into ancilla as q[3] and q[4] of the circuit. A complete state tomography is performed to construct the experimental density matrix for the ancilla.\\
The theoretical density matrix,
\begin{equation}
\begin{split}
\rho^T&=|\Psi\rangle\langle\Psi|\\
&=\begin{pmatrix}
0  &   0  &   0  &   0\\
0  &   0   &  0   &  0\\
0   &  0   &  0   &  0\\
0   &  0  &   0   &  1
\end{pmatrix}
\end{split}
\end{equation}
The experimental density matrix,
\begin{equation*}
\rho^E= Re[\rho^E] + \iota Im[\rho^E]
\end{equation*} 
\begin{equation}
Re[\rho^E]=
\begin{pmatrix}
      0 &  -0.0010  &  0.0060 &  -0.0008\\
-0.0010  &       0  &  0.0008  &  0.0020\\
0.0060  &  0.0008    &     0  & -0.0030\\
-0.0008  &  0.0020  & -0.0030 &   1.0000
\end{pmatrix}
\end{equation}
\begin{equation}
Im[\rho^E]=
\begin{pmatrix}
0  & -0.0003 &  -0.0043 &  -0.0115\\
0.0003      &   0 &  -0.0015  &  0.0057\\
0.0043  &  0.0015     &    0  & -0.0003\\
0.0115  & -0.0057 &   0.0003     &    0
\end{pmatrix}
\end{equation}

\begin{figure}[H]
	\begin{center}
		\includegraphics[scale=0.5]{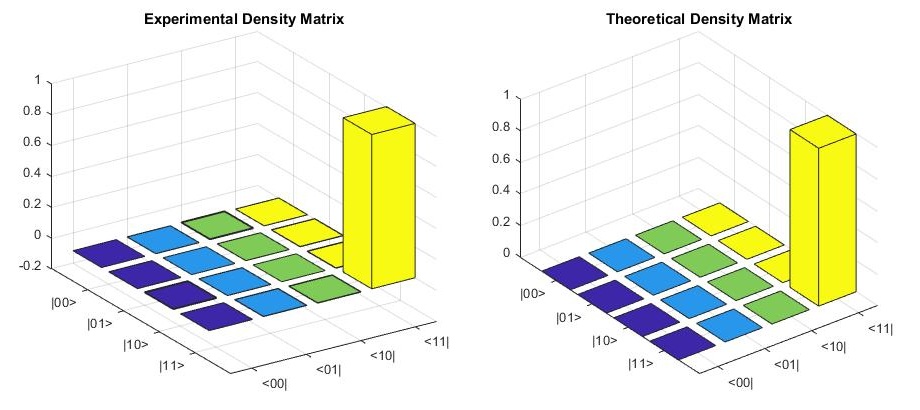}
		\caption{\label{t_parityan}Construction of the state  $|\Psi\rangle$.}
	\end{center}
\end{figure}

Average absolute deviation, $\langle\Delta x\rangle=0.17\% $, maximum absolute deviation, $\Delta x_{max}=0.6\%$ and fidelity, $F(\rho^T,\rho^E)=\sqrt{\langle\Psi|\rho^E|\Psi\rangle}= 1$.
\subsection{Error correction}
\subsubsection{Bell state}
In section-\ref{secbell}, according to the implemented circuit, the final state is $|\Psi\rangle=\frac{1}{2}(|01\rangle-|10\rangle)$. $|\Psi\rangle$ is constructed at $q[0]$ and $q[1]$, which means complete state tomography is imposed at these qubits.\\
The theoretical density matrix,
\begin{equation}
\begin{split}
\rho^T&=|\Psi\rangle\langle\Psi|\\
&=\begin{pmatrix}
0 & 0 & 0 & 0\\
0 & 0.5 & -0.5 & 0\\
0 & -0.5 & 0.5 & 0\\
0 & 0 & 0 & 0
\end{pmatrix}
\end{split}
\end{equation}
The experimental density matrix,
\begin{equation*}
\rho^E= Re[\rho^E] + \iota Im[\rho^E]
\end{equation*} 
\begin{equation}
Re[\rho^E]=
\begin{pmatrix}
0 & -0.0020 & -0.0040 & 0\\
-0.0020 & 0.4900 & -0.5000 & 0.0040\\
-0.0040 & -0.5000 & 0.5100 & 0.0020\\
0 & 0.0040 & 0.0020 & 0\\
\end{pmatrix}
\end{equation}
\begin{equation}
Im[\rho^E]=
\begin{pmatrix}
0 & -0.0035 & 0.0025 & -0.0027\\
0.0035 & 0 & 0.0033 & 0.0055\\
-0.0025 & -0.0033 & 0 & -0.0045\\
0.0027 & -0.0055 & 0.0045 & 0\\
\end{pmatrix}
\end{equation}

\begin{figure}[H]
	\begin{center}
		\includegraphics[scale=0.5]{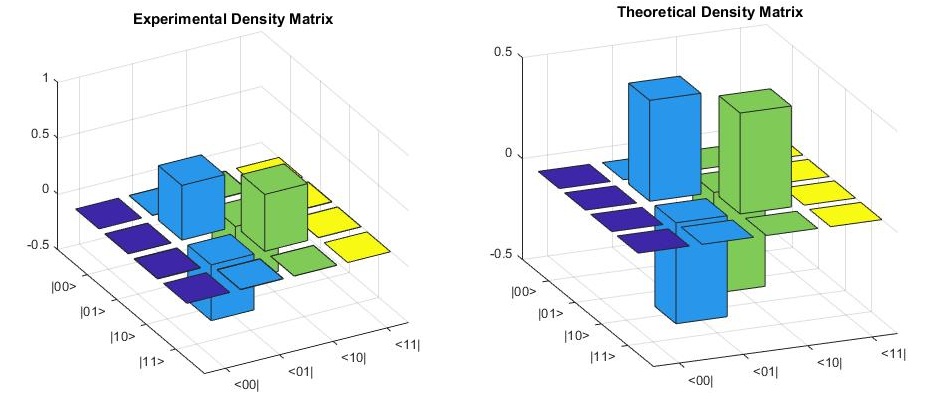}
		\caption{\label{t_bell}Construction of Bell state  $|\Psi\rangle=\frac{1}{2}(|01\rangle-|10\rangle)$}
	\end{center}
\end{figure}

Average absolute deviation, $\langle\Delta x\rangle=0.28\% $, maximum absolute deviation, $\Delta x_{max}=1\%$ and fidelity, $F(\rho^T,\rho^E)=\sqrt{\langle\Psi|\rho^E|\Psi\rangle}= 1$.

\subsubsection{GHZ state}
In section-\ref{seccor}, we have corrected an erroneous state $|\Psi^{e}\rangle$ to a state $|\Psi^{-}_{010}\rangle$ as the initial phase and parity are $|1\rangle$ and $|11\rangle$ respectively. At the end of the circuits i.e. bit-flip correction (section-\ref{secbit}) we will get our desired state $|\Psi^{-}_{010}\rangle$. So a complete state tomography for GHZ state(first three qubits of the circuit as they are encoded for GHZ state) is needed for the verification.\\
The theoretical density matrix,
\begin{equation}
\begin{split}
\rho^T&=|\Psi^{-}_{010}\rangle\langle\Psi^{-}_{010}|\\
&=\begin{pmatrix}
0 & 0 & 0 & 0 & 0 & 0 & 0 & 0\\
0 & 0 & 0 & 0 & 0 & 0 & 0 & 0\\
0 & 0 & 0.5 & 0 & 0 & 0.5 & 0 & 0\\
0 & 0 & 0 & 0 & 0 & 0 & 0 & 0\\
0 & 0 & 0 & 0 & 0 & 0 & 0 & 0\\
0 & 0 & 0.5 & 0 & 0 & 0.5 & 0 & 0\\
0 & 0 & 0 & 0 & 0 & 0 & 0 & 0\\
0 & 0 & 0 & 0 & 0 & 0 & 0 & 0\\
\end{pmatrix}
\end{split}
\end{equation}
The experimental density matrix,
\begin{equation*}
\rho^E= Re[\rho^E] + \iota Im[\rho^E]
\end{equation*} 

\begin{equation}
Re[\rho^E]=
\begin{pmatrix}
    0.0000  &  0.0020 &  -0.0046 &  -0.0017  &  0.0053 &   0.0025 &   0.0017  &       0\\
0.0020  & 0.0000 &  -0.0040  &  0.0006  &  0.0025  &  0.0012      &   0 &  -0.0032\\
-0.0046  & -0.0040  &  0.5010 &   0.0010 &  -0.0020 &  -0.5000  & -0.0035  &  0.0020\\
-0.0017  &  0.0006  &  0.0010  &  0.0000   &      0  &  0.0005   &      0 &   0.0030\\
0.0053  &  0.0025 &  -0.0020    &     0 &  0.0000  &  0.0020  & -0.0051  &  0.0007\\
0.0025 &   0.0012  & -0.5000  &  0.0005 &   0.0020  &  0.4990 &  -0.0010 &   0.0021\\
0.0017    &     0 &  -0.0035   &      0 &  -0.0051  & -0.0010  &  0.0000 &  -0.0020\\
0 &  -0.0032 &   0.0020  &  0.0030 &   0.0007 &   0.0021  & -0.0020 &  0.0000
\end{pmatrix}
\end{equation}
\begin{equation}
Im[\rho^E]=
\begin{pmatrix}
0  & -0.0004 &   0.0013 &  -0.0017  &  0.0011 &   0.0055 &  -0.0004 &   0.0021\\
0.0004     &    0 &   0.0027  &  0.0008 &   0.0055  &  0.0029 &  -0.0021  & -0.0006\\
-0.0013 &  -0.0027    &     0  &  0.0021  &  0.0076  &  0.0016 &  -0.0014 &  -0.0012\\
0.0017  & -0.0008  & -0.0021    &     0  & -0.0006 &  -0.0026 &  -0.0027  & -0.0026\\
-0.0011  & -0.0055  & -0.0076  &  0.0006  &       0  &  0.0011  &  0.0018 &  -0.0020\\
-0.0055  & -0.0029 &  -0.0016  &  0.0026 &  -0.0011  &       0 &  -0.0030  &  0.0023\\
0.0004  &  0.0021  &  0.0014  &  0.0027 &  -0.0018  &  0.0030   &      0 &  -0.0009\\
-0.0021  &  0.0006 &   0.0012  &  0.0026  &  0.0020  & -0.0023  &  0.0009    &     0\\
\end{pmatrix}
\end{equation}
\begin{figure}[H]
	\begin{center}
		\includegraphics[scale=0.6]{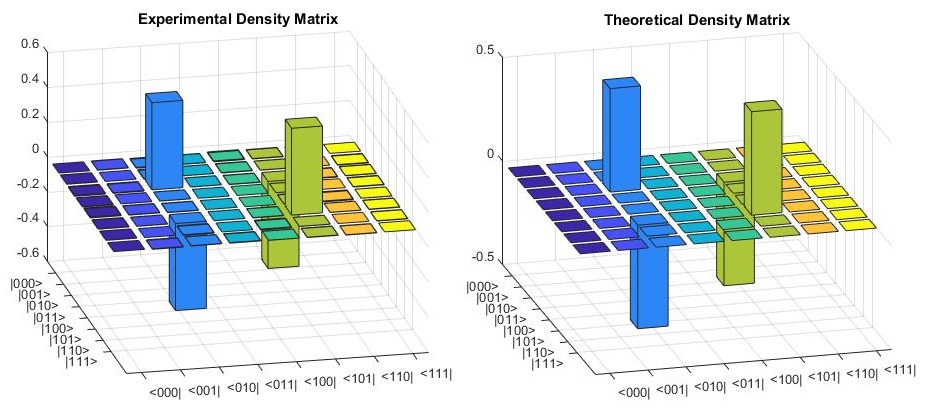}
		\caption{\label{t_parity}Construction of the GHZ state $|\Psi^{-}_{010}\rangle$.}
	\end{center}
\end{figure}

Average absolute deviation, $\langle\Delta x\rangle=0.17\% $, maximum absolute deviation, $\Delta x_{max}=0.53\%$ and fidelity, $F(\rho^T,\rho^E)=\sqrt{\langle\Psi^{-}_{010}|\rho^E|\Psi^{-}_{010}\rangle}= 1$.

\section{Generalization in $\mathbb{C}^{d^n}$} \label{secqudit}

Above automated error correction circuit can be generalized for qudits, based on the fact that the non-destructive discrimination algorithm, which is the backbone of our circuit, is easily extendable to higher dimensions \cite{gupt1}, \cite{Gupt}, \cite{Pani}. Thus by extending our error correcting circuit in the same lines of discriminatory algorithm, will result in the correction of errors in higher dimensions.  

\subsection{Definitions}

Bell State discrimination can be easily generalized to entangled states of n qudits (d-dimensional states). The formalism is adopted from Panigrahi et al.\cite{Pani}.

The Pauli matrices are replaced by their d-dimensional analogs. The X and Z gates are generalized to $X_d$ and $Z_d$ respectively, 

\begin{equation}
Z_{d}|j\rangle = e^{2\pi ij/d}|j\rangle
\end{equation}
\begin{equation}
X_{d}|j\rangle = |j-1\rangle
\end{equation}
\begin{equation}
	X_d^\dagger|j\rangle=|j+1\rangle
\end{equation}
where the change in the ket is in mod $d$ arithmetic. The generalized Hadamard transform given by,\\
\begin{equation}
H_{d}|j\rangle = \frac{1}{\sqrt{d}}\sum_{k=0}^{d-1} e^{2\pi ijk/d}|k\rangle
\end{equation}\\
\begin{equation}
	H_{d}^\dagger|j\rangle = \frac{1}{\sqrt{d}}\sum_{k=0}^{d-1} e^{-2\pi ijk/d}|k\rangle
\end{equation}
such that the operators $X_d$ and $Z_d$ are related as $X_d = H_d Z_d H_d^\dag$.
Note that unlike the qubit case, $Z_d, X_d$ and $H_d$ are not Hermitian. The CNOT gate($C_X$) is generalized to $C_{X_d}$ where,
\begin{equation}
	C_{{X_d}_{1\rightarrow2}}|i\rangle|j\rangle=|j-i\rangle\
\end{equation}
\begin{equation}
C_{{X_d}_{1\rightarrow2}}^\dagger|i\rangle|j\rangle=|j+i\rangle\
\end{equation}
The representation $C_{{X_d}_{1\rightarrow2}}|i\rangle|j\rangle$ means $|i\rangle$ as the control qudit and $|j\rangle$ as the target qudit.\\ 
The generalised maximally entangled states for n-qudits can be represented as,

\begin{equation}
|\Psi_{pq_{1}q_{2}.....q_{n-1}}\rangle = \frac{1}{\sqrt{d}}\sum_{j=0}^{d-1} e^{2\pi ijp/d}|j\rangle|j+q_{1}\rangle|j+q_{2}\rangle......|j+q_{n-1}\rangle
\end{equation}

Here $p$ is an integer representing the phase and $q_{i}$ the parity of the $i+1^{th}$ qudit.

\subsection{Phase measurement}

Clearly, analogous to the two-dimensional case, the phase and parity information of the desired d-dimensional state to be sent, should first be attained on the ancilla qudit\cite{Gupt}. For this purpose, we use the following operations on the composite Hilbert space of the original d-dimensional state and the ancilla.
\begin{equation}
|\Psi_{pq_{1}q_{2}.....q_{n-1}}\rangle |p\rangle = [I^{\otimes n} \otimes H^{\dagger}_{d}] \times [\bigotimes_{m=1}^{m=n} C_{X_d} (|\Psi_m\rangle \longleftarrow |A\rangle)] \times [I^{\otimes n} \otimes H_{d}]|\Psi_{pq_{1}q_{2}.....q_{n-1}}\rangle |0_{A}\rangle
\end{equation}

Clearly, as the equation suggests, the phase bit information gets encoded into the ancilla qudit.

\subsection{Parity measurement}
For the parity measurement, we use the following operations:

\begin{equation}
|\Psi_{pq_{1}q_{2}.....q_{n-1}}\rangle |q_{i} - q_{i-1}\rangle = [C_{X_{d}}^{\dagger} (|\Psi_i\rangle \longrightarrow |A_i\rangle)C_{X_{d}} (|\Psi_{i-1}\rangle \longrightarrow |A_i\rangle)]|\Psi_{pq_{1}q_{2}.....q_{n-1}}\rangle |0_{A_{i}}\rangle
\end{equation}

\begin{figure}[H]
	\begin{center}
		\includegraphics[scale=0.35]{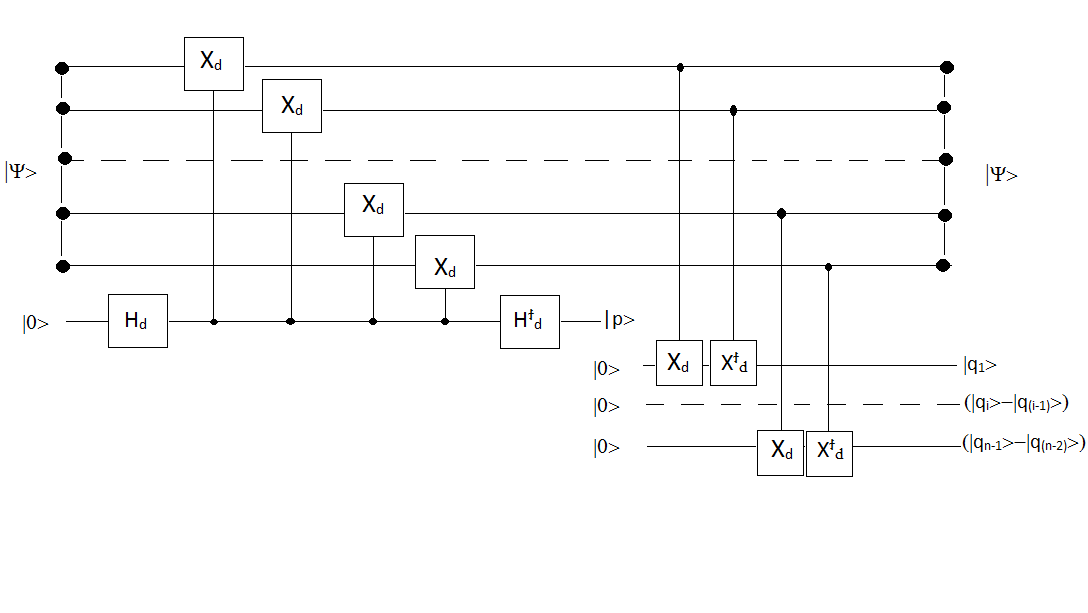}
		\caption{\label{d_check}Phase and parity measurement algorithm. Here the left part is for phase measurement and right part is for parity measurement.}
	\end{center}
\end{figure}

The ancilla qudit can then be measured to attain the relative parity information\cite{Gupt}. Here $i$ runs from $1$ to $n-1$ but when $i=1$ then $q_0=|0\rangle$ and $|\Psi_0\rangle=|p\rangle$ . Note that the operators in the qudit space are not Hermitian unlike in the qubit space. Both parts of the algorithm are clearly depicted in fig-\ref{d_check}.
\subsection{Error correction algorithm}
The error correction is completed in three steps, as in 2-D case. The state to be corrected can be represented as,

\begin{equation}
|\Psi^{e}\rangle = \frac{1}{\sqrt{d}}\sum_{j=0}^{d-1} e^{2\pi ijp^{'}/d}e^{i\delta_{j}}|j\rangle|j+q_{1}^{'}\rangle|j+q_{2}^{'}\rangle......|j+q_{n-1}^{'}\rangle
\end{equation}

\subsubsection{Step 1}

This step deals with the removal of arbitrary phases introduced into the state\cite{auto}. The phase of the erroneous state is transferred to the ancilla. This is a result of the entanglement of original state with the ancilla on application of $C_{X_d}$, followed by disentanglement by $C_{X_d}^{\dagger}$. It can be clearly seen here:

\begin{equation}
|\Psi^{e1}\rangle |t\rangle = [[C_{X_d}^{\dagger} (|\Psi_0\rangle \longrightarrow |A\rangle)] \otimes I^{\otimes n} ] \times [\bigotimes_{m=1}^{m=n} C_{X_d} (|\Psi_m\rangle \longleftarrow |A\rangle)] \times [I^{\otimes n} \otimes H_{d}]|\Psi^{e}\rangle |0_{A}\rangle
\end{equation}

\[= (\frac{1}{\sqrt{d}}\sum_{j=0}^{d-1} e^{2\pi ijp^{'}/d}e^{i\delta_{j}}|j\rangle|j+q_{1}^{'}\rangle|j+q_{2}^{'}\rangle......|j+q_{n-1}^{'}\rangle) [\sum_{k=0}^{d-1} |k\rangle]\]

\[= \frac{1}{\sqrt{d}}\sum_{j,k=0}^{d-1} e^{2\pi i(j+k)p^{'}/d}e^{i\delta_{j+k}}|j\rangle|j+q_{1}^{'}\rangle|j+q_{2}^{'}\rangle......|j+q_{n-1}^{'}\rangle |k\rangle\]

\[= \frac{1}{\sqrt{d}}\sum_{j,k=0}^{d-1} e^{2\pi i(j+k)p^{'}/d}e^{i\delta_{j+k}}|j\rangle|j+q_{1}^{'}\rangle|j+q_{2}^{'}\rangle......|j+q_{n-1}^{'}\rangle |k+j\rangle\]

\[= \frac{1}{\sqrt{d}}\sum_{j,k=0}^{d-1} e^{2\pi ikp^{'}/d}e^{i\delta_{k}}|j\rangle|j+q_{1}^{'}\rangle|j+q_{2}^{'}\rangle......|j+q_{n-1}^{'}\rangle |k\rangle\]

\begin{equation}
= ( \frac{1}{\sqrt{d}}\sum_{j=0}^{d-1}|j\rangle|j+q_{1}^{'}\rangle|j+q_{2}^{'}\rangle......|j+q_{n-1}^{'}\rangle) [\sum_{k=0}^{d-1} e^{2\pi ikp^{'}/d}e^{i\delta_{k}}|k\rangle]
\end{equation}

\subsubsection{Step 2- Phase difference}

If we are interested in measuring the phase error in the state, we can use the following operation:

\begin{equation}
\begin{split}
|\Psi_{p}\rangle |p-p^{'}\rangle &= [I^{\otimes n} \otimes H_{d}] \times[C_{X_d}^{\dagger} (|\Psi_0\rangle \longrightarrow |A\rangle)] \otimes I^{\otimes n} ] \times [\bigotimes_{m=1}^{m=n} C_{X_d} (|\Psi_m\rangle \longleftarrow |A\rangle)]\\&\hspace{1cm}\times [I^{\otimes n} \otimes H^{\dagger}_{d}]|\Psi_{p^{'}q_{1}^{'}q_{2}^{'}.....q_{n-1}^{'}}\rangle |p\rangle
\end{split}
\end{equation}

\[
=( \frac{1}{\sqrt{d}}\sum_{j=0}^{d-1} e^{2\pi ijp^{'}/d}|j\rangle|j+q_{1}^{'}\rangle|j+q_{2}^{'}\rangle......|j+q_{n-1}^{'}\rangle) [\sum_{k=0}^{d-1} e^{-2\pi ikp/d} |k\rangle]
\]

\[
=\frac{1}{\sqrt{d}}\sum_{j,k=0}^{d-1} e^{2\pi ijp^{'}/d}e^{2\pi ik(p^{'}-p)/d}|j\rangle|j+q_{1}^{'}\rangle|j+q_{2}^{'}\rangle......|j+q_{n-1}^{'}\rangle |k\rangle
\]

\[
=\frac{1}{\sqrt{d}}\sum_{j,k=0}^{d-1} e^{2\pi ijp/d}e^{2\pi ik(p^{'}-p)/d}|j\rangle|j+q_{1}^{'}\rangle|j+q_{2}^{'}\rangle......|j+q_{n-1}^{'}\rangle |k\rangle
\]

\begin{equation}
=( \frac{1}{\sqrt{d}}\sum_{j=0}^{d-1} e^{2\pi ijp/d}|j\rangle|j+q_{1}^{'}\rangle|j+q_{2}^{'}\rangle......|j+q_{n-1}^{'}\rangle) [|p - p^{'}\rangle]
\end{equation}

\subsubsection{Step 2- Phase correction}

\begin{figure}[H]
	\begin{center}
		\includegraphics[scale=0.4]{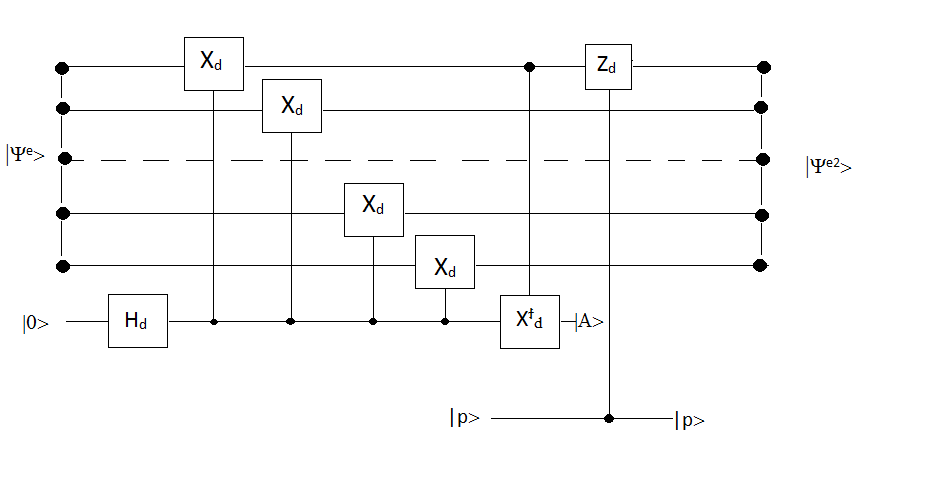}
		\caption{\label{d_phase}Phase error correction. Here phase is corrected by following step 1 and step 2 (phase correction) algorithms.}
	\end{center}
\end{figure}

Note that the step given above is redundant after the state goes through step 1, if we're not interested in getting the phase error information on the ancilla. The following operation is sufficient for the phase correction:

\begin{equation}
|\Psi^{e2}\rangle|p \rangle = [C_{Z_d}(|\Psi_0\rangle \longleftarrow |p\rangle)] |\Psi^{e1} \rangle |p \rangle
\end{equation}
\begin{equation}
=( \frac{1}{\sqrt{d}}\sum_{j=0}^{d-1} e^{2\pi ijp/d}|j\rangle|j+q_{1}^{'}\rangle|j+q_{2}^{'}\rangle......|j+q_{n-1}^{'}\rangle) [|p\rangle]
\end{equation}
where, $C_{Z_d}(|\Psi_0\rangle \longleftarrow |p\rangle)=(Z_d)^{p}|\Psi_{0}\rangle$. Step 1 and 2 are illustrated in fig-\ref{d_phase}.
\subsubsection{Step 3}
\begin{figure}[H]
	\begin{center}
		\includegraphics[scale=0.5]{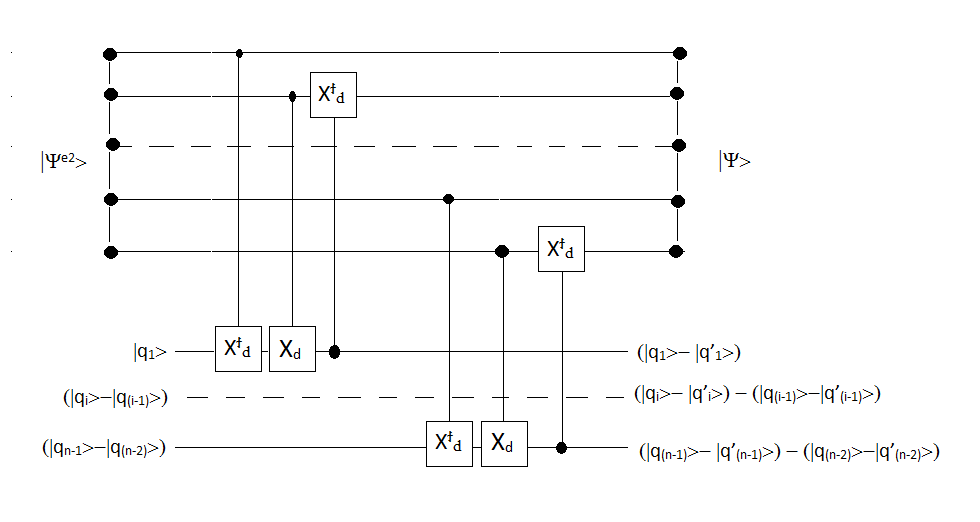}
		\caption{\label{d_parity}Parity correction.}
	\end{center}
\end{figure}

This is the final step required to retrieve the original $n$-qudit state. This operation is based on the same principle as step 1. For $n$-qudit system, $n-1$ ancilla are required. Similar to the 2-d case, we can retrieve the original qudit sequence by using the relative parity information obtained from the original state.

\begin{equation}
\begin{split}
|\Psi_{pq_{1}q_{2}.....q_{n-1}}\rangle |q_{i} - q_{i}^{'}\rangle &= [C_{X_{d}}^{\dagger} (|\Psi_i\rangle \longleftarrow |A_i\rangle)]\times[  C_{X_{d}} (|\Psi_i\rangle \longrightarrow |A_i\rangle)]\\&\hspace{1cm}\times[C_{X_{d}}^{\dagger} (|\Psi_{i-1}\rangle \longrightarrow |A_i\rangle)]|\Psi_{pq_{1}^{'}q_{2}^{'}.....q_{n-1}^{'}}\rangle|q_i-q_{i-1}\rangle
\end{split}
\end{equation}

\[
=( \frac{1}{\sqrt{d}}\sum_{j=0}^{d-1} e^{2\pi ijp/d}|j\rangle|j+q_{1}^{'}\rangle|j+q_{2}^{'}\rangle......|j+q_{n-1}^{'}\rangle) (|q_{i} - q_{i-1}\rangle)
\]

\[
=\frac{1}{\sqrt{d}}\sum_{j=0}^{d-1} e^{2\pi ijp/d}|j\rangle|j+q_{1}^{'}\rangle|j+q_{2}^{'}\rangle......|j+q_{n-1}^{'}\rangle |(q_{i} - q_{i}^{'}) - (q_{i-1} - q_{i-1}^{'})\rangle
\]

\begin{equation}
=(\frac{1}{\sqrt{d}}\sum_{j=0}^{d-1} e^{2\pi ijp/d}|j\rangle|j+q_{1}\rangle|j+q_{2}\rangle......|j+q_{i}...\rangle )[|(q_{i} - q_{i}^{'}) - (q_{i-1} - q_{i-1}^{'})\rangle]
\end{equation} 
Here $q_{i-1}=q_{i-1}^{'}$ as this algorithm corrected parity upto i-1 and now it is correcting for i. So the final ancilla will be $|q_i-q_i^{'}\rangle$. Step 3 is illustrated in fig-\ref{d_parity}.

The original state is retrieved as given in equation (38) above.

\section{Conclusion}\label{seccon}
Experimental verification of automated error correction algorithm for Bell states and GHZ states, and nondestructive discrimination of GHZ states have been tested on a five-qubit quantum computer. Due to the limitation of IBM quantum computer\cite{exp}, GHZ states' discrimination and correction have been performed in different stages. Our experimental results are consistent with the theoretical results given by Pandey et al.\cite{auto}. Quantum state tomography results show retainment of the Generalized Bell State's identity. It confirms with the average absolute deviation of $0.05\% \sim 0.28\%$ , maximum absolute deviation of $0.1\% \sim 1\%$ and fidelity of being 1 for all tomography results. The advantage of the use of ancilla based approach, is that it can be carried out with minimal interference to the quantum circuit of which the above entangled state forms a part. Subsequently, the procedure for automated error correction for the generalized entangled qudit state is explicated whose experimental implementation requires more involved phase gates. We hope this qudit entangled state discrimination and error correction finds experimental verification in near future. We aim that the present experiment will soon be extended to the experimental discrimination and automated error correction of more complex entangled states.  
\section{Acknowledgments}
DG and BKB are financially supported by DST Inspire Fellowship. PA would like to thank the National Initiative for Undergraduate Science (NIUS) Physics. The authors are extremely grateful to IBM team and IBM Quantum Experience project. This work does not reflect the views or opinions of IBM or any of its employees.

\newpage 
\begin{center}
\appendix
\hypertarget{ap1}{\textbf{Appendix 1: Results of all above circuits in IBM quantum computer}}
\begin{figure}[H]
	\begin{center}
		\subfloat[Correction of Bell state]{
			\includegraphics[scale=0.5]{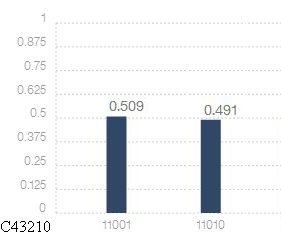}}
		\hspace{3.8cm}
		\subfloat[Phase-checking of $|\Psi_{010}^{-}\rangle$]{
			\includegraphics[scale=0.5]{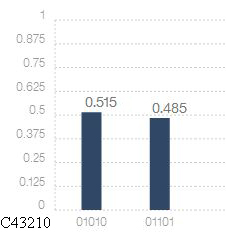}}
	\end{center}
\end{figure}
\begin{figure}[H]
	\begin{center}	
		\subfloat[Parity-checking of $|\Psi_{010}^{-}\rangle$]{
			\includegraphics[scale=0.5]{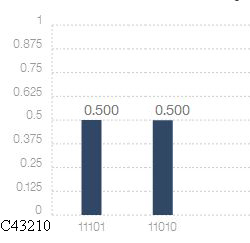}}
		\hspace{3.8cm}
		\subfloat[Arbitrarily phase correction]{
			\includegraphics[scale=0.5]{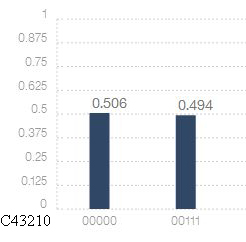}}
	\end{center}
\end{figure}
\begin{figure}[H]
	\begin{center}		
		\subfloat[Phase flip correction when $|\phi\rangle=|1\rangle$]{
			\includegraphics[scale=0.5]{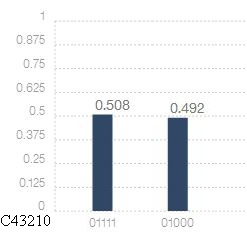}}
		\hspace{4cm}
		\subfloat[Bit flip correction when $|p\rangle=|11\rangle$]{
			\includegraphics[scale=0.5]{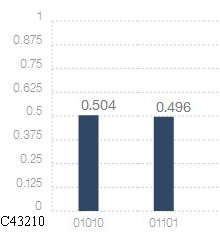}}
	\end{center}
\end{figure}
\newpage
\appendix
\hypertarget{ap2}{\textbf{Appendix 2: Calibration data of the IBM quantum computer}}
\begin{figure}[H]
	\includegraphics[width=16cm]{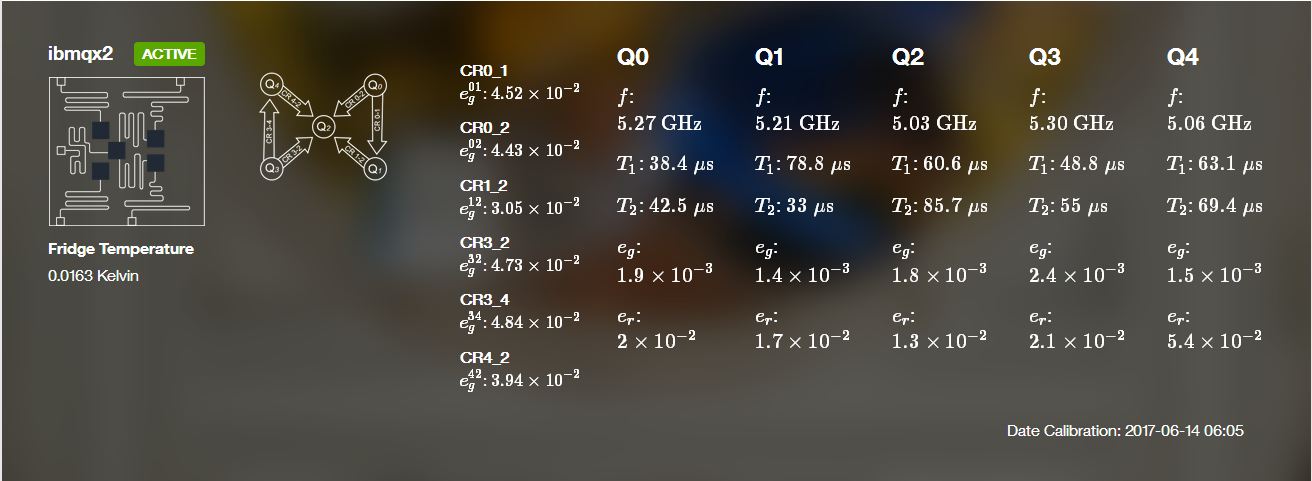}
\end{figure}
\end{center}


\begin{thebibliography}{100}
	
	\bibitem{nil}M.A. Nielsen and I.L. Chuang, Quantum Computation and Quantum Information, Cambridge University Press, \href{http://dl.acm.org/citation.cfm?id=544199}{ISBN:0-521-63503-9 (2000)}.
	\bibitem{ben}C.H. Bennett, G. Brassard, C. Crepeau, R. Jozsa, A. Peres, and W.K. Wootters, \href{https://doi.org/10.1103/PhysRevLett.70.1895}{Phys. Rev. Lett. 70 1895 (1993)}.
	\bibitem{tele}S. Ghosh, G. Kar, A. Roy, D. Sarkar, and U. Sen, \href{https://doi.org/10.1088/1367-2630/4/1/348}{New J. Phys. 4, 48 (2002)}.
	\bibitem{sre1}S. Muralidharan and P.K. Panigrahi, \href{https://doi.org/10.1103/PhysRevA.77.032321}{Phys. Rev. A \textbf{77}, 032321 (2008)}.
    \bibitem{sre3}S. Choudhury, S. Muralidharan, and P.K. Panigrahi, \href{https://doi.org/10.1088/1751-8113/42/11/115303}{J. Phys. A: Math. Theor., \textbf{42} 115303 (2009)}.
    \bibitem{sre7}S. Muralidharan, S. Karumanchi, S. Jain, R. Srikanth, and P.K. Panigrahi, \href{https://doi.org/10.1140/epjd/e2010-09653-x}{Eur. Phys. J. D \textbf{61}, 757-763, 2011}
	\bibitem{sre9}N. Paul, J.V. Menon, S. Karumanchi, S. Muralidharan, and P.K. Panigrahi, \href{\doibase 10.1007/s11128-010-0217-7}{Quantum Inf. Process., \textbf{10}(5), 619-632 (2011)}
	\bibitem{sre2}S. Muralidharan and P.K. Panigrahi, \href{https://doi.org/10.1103/PhysRevA.78.062333}{Physical Review A \textbf{78}(6), 062333 (2008)}.
	\bibitem{sre4}S. Muralidharan, S. Karumanchi, S. Narayanaswamy, R. Srikanth, and P.K. Panigrahi, \href{https://arxiv.org/abs/0907.3532}{arXiv:quant-ph/0907.3532v2 (2011)}.
	\bibitem{sre5}P.K. Panigrahi, S. Karumanchi, and S. Muralidharan, \href{http://www.ias.ac.in/article/fulltext/pram/073/03/0499-0504}{Pramana - J. Phys., \textbf{73}(3), 499-504 (2009)}.
	\bibitem{sre8}S. Muralidharan, S. Jain and P.K. Panigrahi, \href{https://doi.org/10.1016/j.optcom.2010.10.026}{Optics Communications, \textbf{284}(4), 1082-1085 (2010)}.
	\bibitem{nandi}K. Nandi and G. Paul, \href{https://arxiv.org/abs/1501.07529}{arXiv:quant-ph/1501.07529v2 (2015)}.
    \bibitem{sre6}S. Jain, S. Muralidharan, and P.K. Panigrahi, \href{https://doi.org/10.1209/0295-5075/87/60008}{EPL, \textbf{87}(6) (2009)}. 
	\bibitem{sre10}E.S. Prasath, S. Muralidharan, C. Mitra, and P.K. Panigrahi, \href{\doibase 10.1007/s11128-011-0252-z}{Quantum Inf. Process., \textbf{11}(2), 397-410 (2012)}
    \bibitem{sup}P. Agrawal and A. Pati, \href{https://doi.org/10.1103/PhysRevA.74.062320}{Phys. Rev. A, \textbf{74}, 062320 (2006)}.
	\bibitem{srm}S.R. Moulick and P.K. Panigrahi, \href{\doibase 10.1007/s11128-016-1273-4}{Quantum Inf. Process., \textbf{15}(6), 2475–2486 (2016)}
	\bibitem{bk1}B.K. Behera, A. Banerjee and P.K. Panigrahi, \href{https://arxiv.org/abs/1707.00182}{arXiv:quant-ph/1707.00182v2 (2017)}.
	\bibitem{dia}Y. Xia, C.B. Fu, S. Zhang, S.K. Hong and K.H. Yeon, \href{https://arxiv.org/abs/quant-ph/0601127}{arXiv:quant-ph/0601127v1 (2006)}.
	\bibitem{cor1}D.G. Cory, M.D. Price, W. Maas, E. Knill, R. Laflamme, W.H. Zurek, T.F. Havel and S.S. Somaroo, \href{https://doi.org/10.1103/PhysRevLett.81.2152}{Phys. Rev. Lett. \textbf{81}, 2152 (1998)}.
	\bibitem{cor2}E. Knill, R. Laflamme, G.J. Milburn, \href{https://doi.org/10.1038/35051009}{Nature \textbf{409}, 46-52 (2001)}.
	\bibitem{cor3}J. Chiaverini, D. Leibfried, T. Schaetz, M.D. Barrett, R.B. Blakestad, J. Britton, W.M. Itano, J.D. Jost, E. Knill, C. Langer, R. Ozeri and D.J. Wineland, \href{https://doi.org/10.1038/nature03074}{Nature 432, 602-605 (2004)}. 
	\bibitem{cor4}P. Schindler, J.T. Barreiro, T. Monz, V. Nebendahl, D. Nigg, M. Chwalla, M. Hennrich and R. Blatt, \href{https://doi.org/10.1126/science.1203329}{Science 332, 1059-1061 (2011)}.
	\bibitem{gupt1}M. Gupta, A. Pathak, R. Srikanth, and P.K. Panigrahi, \href{https://arxiv.org/abs/quant-ph/0507096}{arXiv:quant-ph/0507096v1 (2005)}. 
    \bibitem{Gupt}M. Gupta, A. Pathak, R. Srikanth, and P.K. Panigrahi, \href{http://www.worldscientific.com/doi/abs/10.1142/S0219749907003092}{IJQI, \textbf{5}, 62 (2007)}.	
	\bibitem{exp}M. Sisodia, A. Shukla, and A. Pathak, \href{https://arxiv.org/abs/1705.00670}{arXiv:quant-ph/1705.00670v2 (2017)}.
    \bibitem{auto}P. Pandey, S. Prasath, M. Gupta, and P.K. Panigrahi,  \href{https://arxiv.org/abs/1008.2129}{arXiv:quant-ph/1008.2129v2 (2010)}.
    \bibitem{Pani}P.K. Panigrahi, M. Gupta, A. Pathak, and R. Srikanth, \href{http://adsabs.harvard.edu/abs/2006AIPC..864..197P}{AIP, \textbf{864}, 197 (2006)}.
	\bibitem{qe}IBM Quantum Experience, URL: \href{http://www.research.ibm.com/quantum/}{http://www.research.ibm.com/quantum/}.
	\bibitem{Jhar}J.R. Samal, M. Gupta, P.K. Panigrahi, and A. Kumar, \href{http://iopscience.iop.org/article/10.1088/0953-4075/43/9/095508/meta}{J. Phys. B: At. Mol. Opt. Phys. \textbf{43}, 095508 (2010)}.
	\bibitem{anil}V.S. Manu and A. Kumar, \href{https://arxiv.org/abs/1105.2186}{arXiv:quant-ph/1105.2186v1 (2011)}.
	\bibitem{IBM1}M. Sisodia, V. Verma, K. Thapliyal, and A. Pathak, \href{\doibase 10.1007/s11128-017-1526-x}{Quantum Inf. Process (2017)}.
	\bibitem{IBM2}M. Sisodia, A. Shukla, K. Thapliyal, and A. Pathak, \href{https://arxiv.org/abs/1704.05294}{arXiv:quant-ph/1704.05294v1 (2017)}.
	\bibitem{IBM4}A. Majumder, S. Mohapatra, and A. Kumar, \href{https://arxiv.org/abs/1707.07460}{arXiv:quant-ph/1707.07460v1 (2017)}.
	\bibitem{IBM5}A.R. Kalra, S. Prakash, B.K. Behera, and P.K. Panigrahi, \href{https://arxiv.org/abs/1707.09462}{arXiv:quant-ph/1707.09462v1 (2017)}.
	\bibitem{user}IBM quantum computing platform (user guide), URL: \href{https://quantumexperience.ng.bluemix.net/qx/user-guide}{https://quantumexperience.ng.bluemix.net/qx/user-guide}.	
    \bibitem{14}D. Alsina and J.I. Latorre, \href{https://journals.aps.org/pra/abstract/10.1103/PhysRevA.94.012314}{Phys. Rev. A \textbf{94}, 012314 (2016)}.
    \bibitem{26}I.L. Chuang, N. Gershenfeld, M.G. Kubinec, and D.W. Leung, \href{https://doi.org/10.1098/rspa.1998.0170}{The Royal Society, \textbf{454}, 447-467 (1998)}.
    \bibitem{27}D.F. James, P.G. Kwiat, W.J. Munro, and A.G. White, \href{https://journals.aps.org/pra/abstract/10.1103/PhysRevA.64.052312}{Phys. Rev. A \textbf{64}, 052312 (2001)}.
    \bibitem{28}M. Hebenstreit, D. Alsina, J. Latorre, and B. Kraus, \href{\doibase 10.1103/PhysRevA.95.052339}{Phys. Rev. A \textbf{95}, 052339 (2017)}.
    \bibitem{29}R. Rundle, T. Tilma, J. Samson, and M. Everitt, \href{https://journals.aps.org/pra/accepted/33078K24I9f1b001e26d1b452e99051af2827d1ce}{Phys. Rev. A (2017)}.
    \bibitem{30}S. Filipp, P. Maurer, P. Leek, M. Baur, R. Bianchetti, J. Fink, M. Göppl, L. Steffen, J. Gambetta, and A. Blais, \href{https://journals.aps.org/prl/abstract/10.1103/PhysRevLett.102.200402}{Phys. Rev. Lett. \textbf{102}, 200402 (2009)}.
    \bibitem{31}A. Shukla, K. Rao, and T. Mahesh, \href{https://arxiv.org/abs/1304.5851}{arXiv:quant-ph/1304.5851 (2013)}.
    \bibitem{32}A. Pathak, Elements of quantum computation and quantum communication, Taylor \& Francis, \href{http://dl.acm.org/citation.cfm?id=2505466}{ISBN:1466517913 9781466517912 (2013)}.
\end{thebibliography}
\end{document}